\documentclass[%
 reprint,
 amsmath,amssymb,
 aps
]{revtex4-2}

\usepackage{dcolumn}
\usepackage{bm}
\usepackage{caption}
\usepackage{subcaption}
\usepackage{url}
\usepackage{graphicx}
\usepackage{float}
\usepackage{array}
\usepackage{natbib}

\begin{document}

\title{Transient Classification in low SNR Gravitational Wave data using Deep Learning}
\author{Rahul Nigam}
\author{Amit Mishra}%
\author{Pranath Reddy}
\affiliation{Department of Physics, Birla Institute of Technology \& Science Pilani - Hyderabad Campus, Hyderabad, India}

\begin{abstract} 
The recent advances in Gravitational-wave astronomy have greatly accelerated the study of Multi-messenger astrophysics. There is a need for the development of fast and efficient algorithms to detect non-astrophysical transients and noises due to the rate and scale at which the data is being provided by LIGO and other gravitational wave observatories. These transients and noises can interfere with the study of gravitational waves and binary mergers and induce false positives. Here, we propose the use of deep learning algorithms to detect and classify these transient signals. Traditional statistical methods are not well designed for dealing with temporal signals but supervised deep learning techniques such as RNN-LSTM and deep CNN have proven to be effective for solving problems such as time-series forecasting and time-series classification. We also use unsupervised models such as Total variation, Principal Component Analysis, Support Vector Machine,  Wavelet decomposition or Random Forests for feature extraction and noise reduction and then study the results obtained by RNN-LSTM and deep CNN for classifying the transients in low-SNR signals. We compare the results obtained by the combination of various unsupervised models and supervised models. This method can be extended to real-time detection of transients and merger signals using deep-learning optimized GPU's for early prediction and study of various astronomical events. We will also explore and compare other machine learning models such as MLP,  Stacked Autoencoder, Random forests, extreme learning machine, Support Vector machine and logistic regression classifier.
\end{abstract}

\maketitle

\section{Introduction}
Detecting gravitational waves is a huge challenge due to the scales at which these events are experienced by us. LIGO detectors detect space compression as small as 10-21m and to do this the instruments need to be very accurate. However, noise and signals due to non-astrophysical sources can creep into the readings giving false positives. Hence, classification of different signals is crucial in case of detecting gravitational waves. 

In recent times, machine learning and neural networks has proven to be an extremely helpful tool for analyzing large amounts of raw data and infer some practical results from it. Here we use neural networks to classify a transient dataset with low SNR ( signal-to-noise ratio ) of gravitational waves to better detect them.

\section{Methodology}
The dataset used for training our transient detection model is created using eight different classes of transients, namely, Sine Gaussian, Ring Down, Gaussian, Supernova, Cusp, Black Hole Merger, Chirping Sine Gaussian and Blip. The transients are then whitened to reflect the actual LIGO data and also to reduce the effect of low frequency noise in the dataset. We will be working with low SNR signals with the value ranging from 5 to 25.
Below we list the different signals we would be working with.

\subsection{Gaussian(GN)}
Few  non-astrophysical signals are simply  modelled as a gaussian with \( \tau \) values of  0.0005, 0.001, 0.0025, 0.005, 0.0075, 0.01, 0.02 and 0.05.
\begin{equation}
    s(t) = e^{-\frac{(t-t_{0})^{2}}{\tau^{2}}}
\end{equation}

\subsection{Sine-Gaussian(SG)}
These model non-astrophysical glitches that are  significant in the analysis of coalescing compact binaries. \( \tau \) is set to \( 2/f_{0} \) where \( f_{0} \) is the central frequency being varied logarithmically from 100Hz to 2000Hz. 
\begin{equation}
    s(t) = e^{-\frac{(t-t_{0})^{2}}{\tau^{2}}}\sin (2\pi f_{o}(t-t_{o}))
\end{equation}

\subsection{Ringdown(RG)}
These  signals can be characterized by their shorter bandwidth but  longer duration  and are basically sinusoids which are damped. They originate from the quasi-normal modes of a final black hole that has been formed due to coalescing compact binaries. In this case, we set \( \tau = 4/f_{0} \) with \( f_{0} \) being similar to the value used for Sine-Gaussian data set.
\begin{equation}
    s(t) = \begin{cases}
     e^{-\frac{(t-t_{0})}{\tau}}\cos (2\pi f_{o}(t-t_{o}))& \text{ if } t\geq t_{o} \\ 
     0 & \text{ if } t< t_{0}
    \end{cases}
\end{equation}

\subsection{Chirping Sine Gaussian (CSG)}
This signal is similar to sine gaussian, CSG has an additional factor for chirping. These accurately model the white glitches that are frequently observed in LIGO data. The parameter are varied as follows: \( f_{o} \): \{5,100\}, \( \alpha \): \{10,100\} and \( \tau \): \{0.001,0.025\}.
\begin{equation}
    s(t) = \frac{exp(-\frac{(1-i\alpha )(t-t_{o})^{2}}{4\tau^{2}} + 2\pi i(t-t_{o})f_{o})}{(2\pi \tau^{2})^{1/4}}
\end{equation}

\subsection{Supernova (SN)}
One of supernova waveforms namely, the Zwerger-Mueller waveforms are generated by axi-symmetric core collapse of supernovae. These are simulated by hydrodynamical simulations of stellar core collapse by varying the initial conditions like adiabatic index, spin, and differential rotation profile. 78 models are included which consists of a simple analytic equation of state.

\subsection{Cusp(CSP)}
Cosmic strings are generated due to symmetry breaking phase transitions in the early universe, they are modelled as the following cusp like signals:
\begin{equation}
    h(f) = A(f)f^{-4/3}
\end{equation}
These are generated with an exponential roll off after a set value of cut-off frequency  \( f_{o} \) varied between 50 Hz and 2000Hz.

\subsection{Black Hole Merger (LBM) }
Using the Lazarus approach, these waveforms capture the coalescence radiation emitted from a merger of binary black-hole systems. To construct the time domain which replicates the merger scenarios, an analytical approximation is used. This is done considering the black hole binaries have a chirp mass in the range \{20,50\} and cos of inclination angle varied
between zero and one.

\subsection{Blip (Blip) }
Although frequently observed in LIGO detectors, the origin of blip signals are not well understood. Here, we simulate these by clipping sine-gaussian at a small percentage level around mean amplitude.

Below we present some generated signals of the different types listed above which will further be used to train our model. In total, we plan to generate 48,000 ( 6000 in each class ) transient signals for training and testing.
\begin{figure}[H]
        \centering
        \begin{subfigure}{\linewidth}
            \centering
            \includegraphics[width=3.5in]{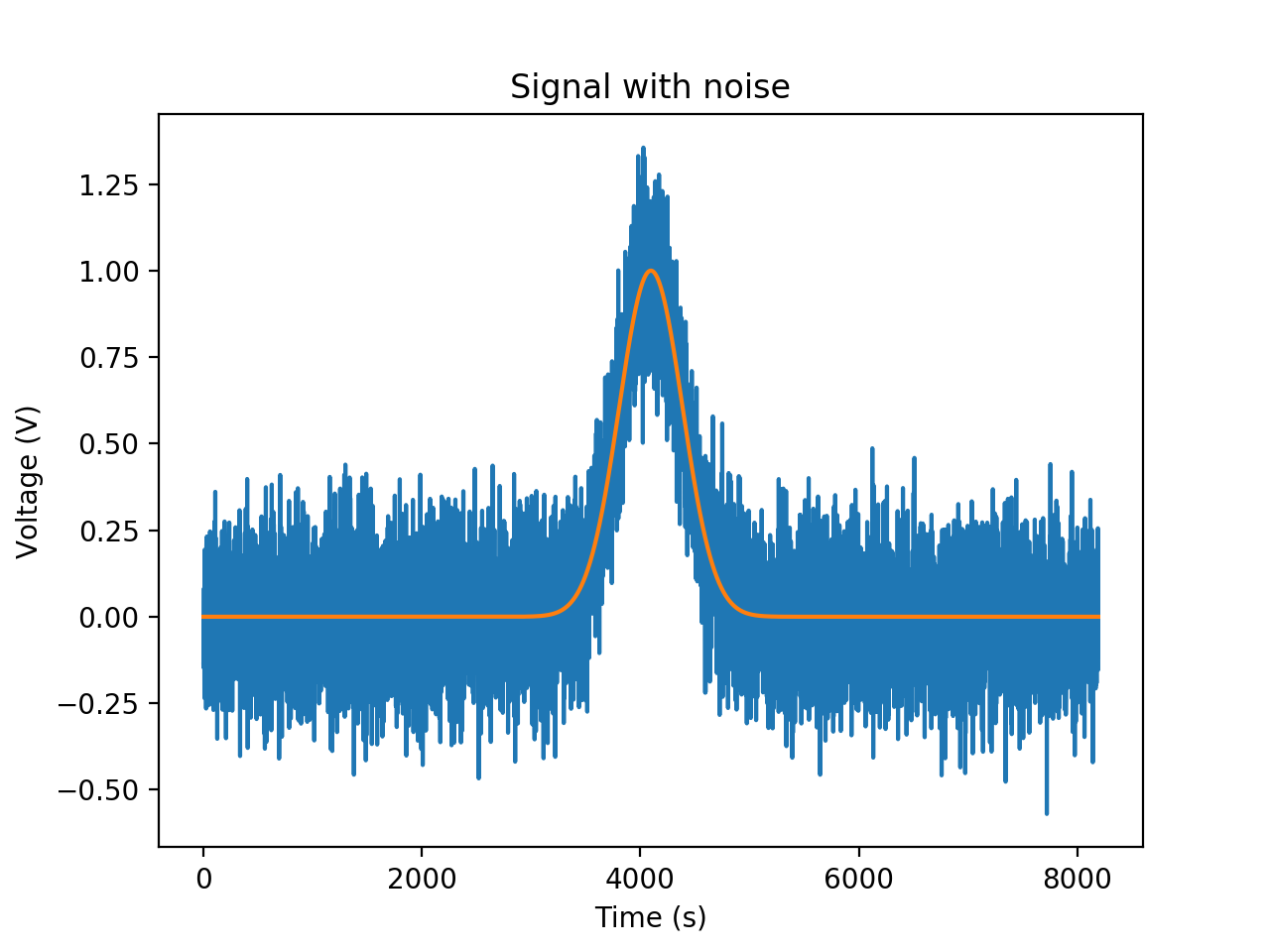}
            \caption{Gaussian}
        \end{subfigure}
        \hfill
        \begin{subfigure}{\linewidth}
            \centering
            \includegraphics[width=3.5in]{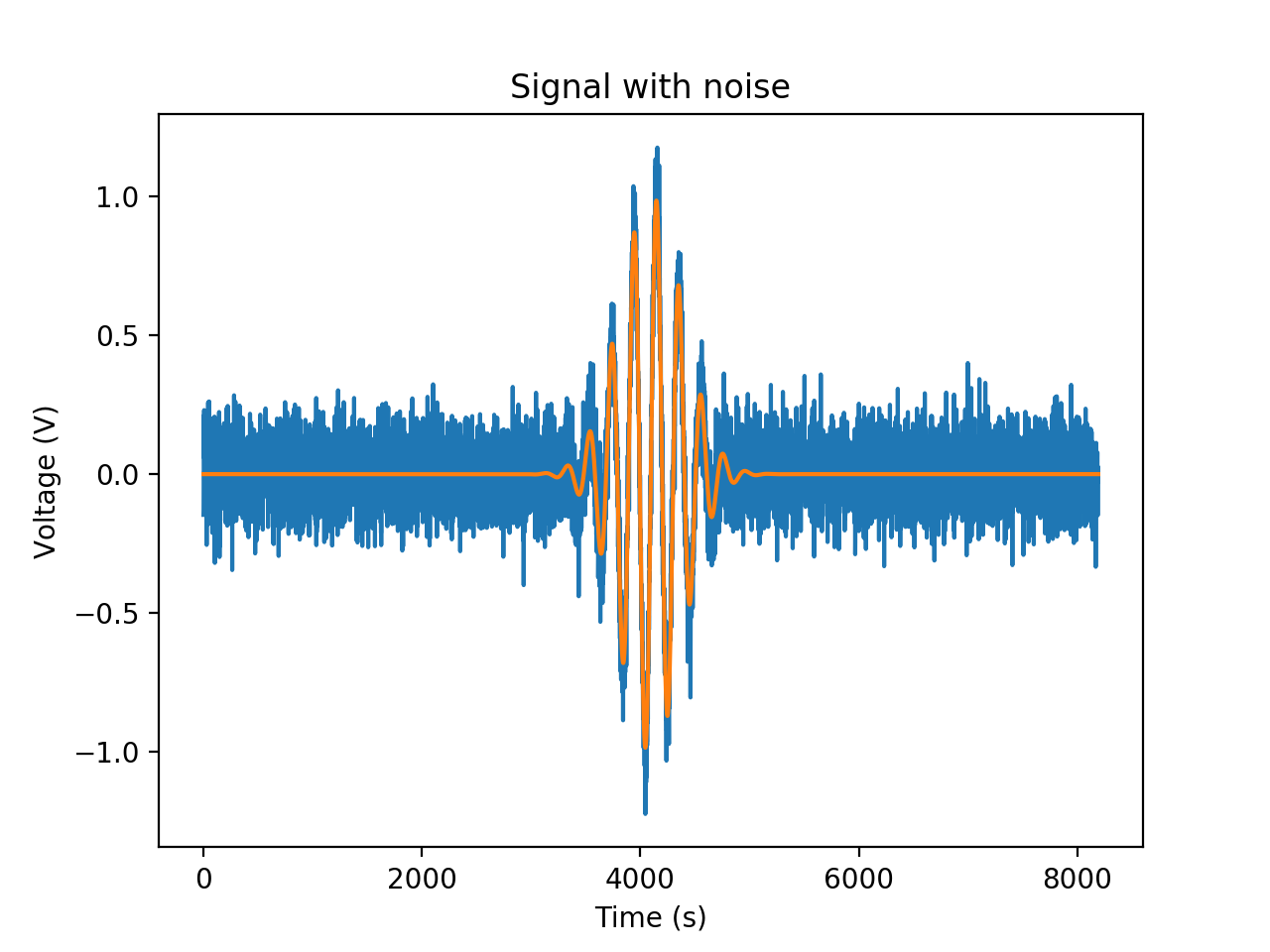}
            \caption{Sine-Gaussian}
            \begin{minipage}{.1cm}
            \vfill
            \end{minipage}
        \end{subfigure} 
        \begin{subfigure}{\linewidth}
            \centering
            \includegraphics[width = 3.5in]{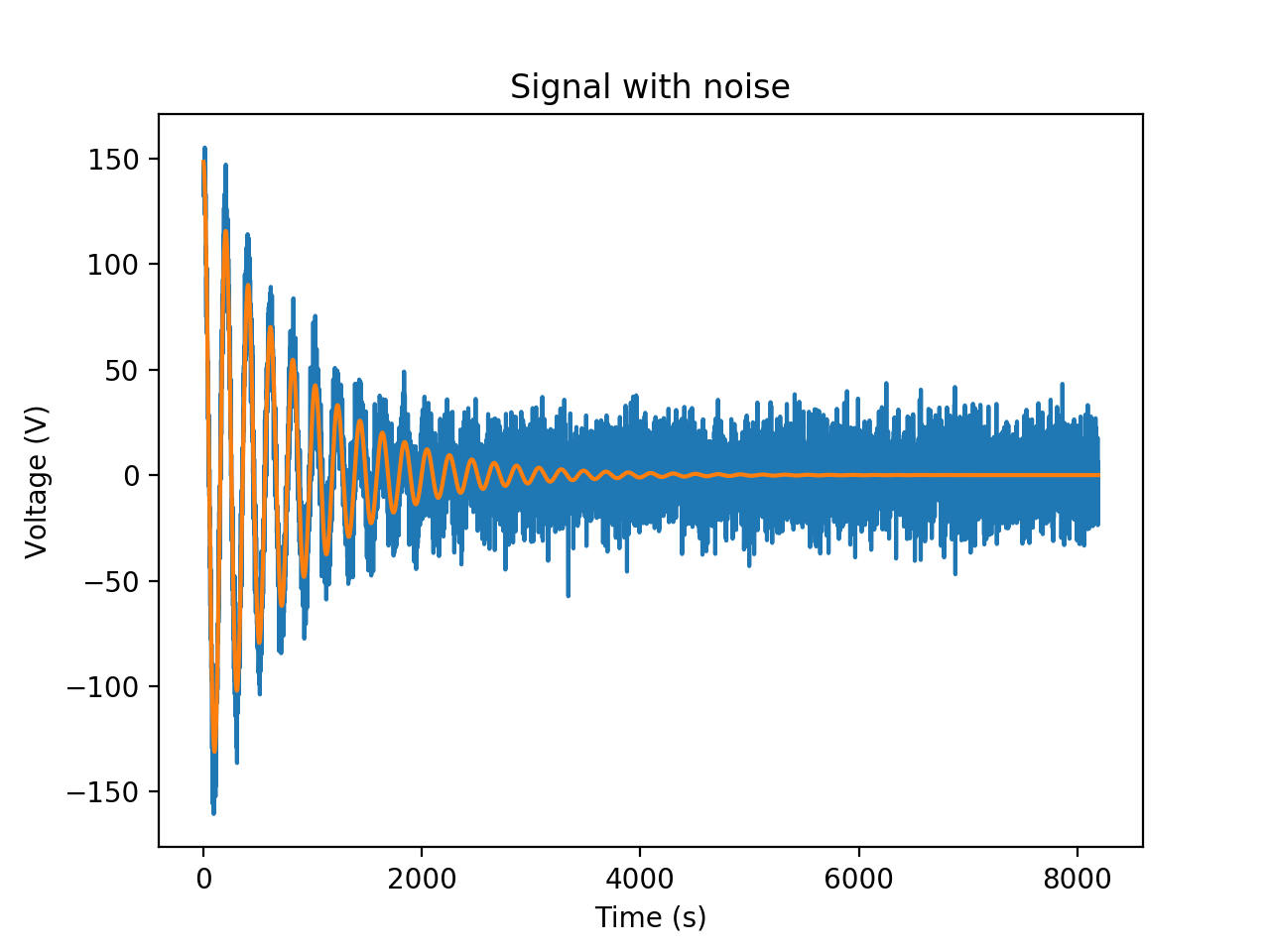}
            \caption{Ringdown}
        \end{subfigure}
        \phantomcaption
\end{figure}
\begin{figure}[H]
        \ContinuedFloat
        \begin{subfigure}{\linewidth}
            \centering
            \includegraphics[width=3.5in]{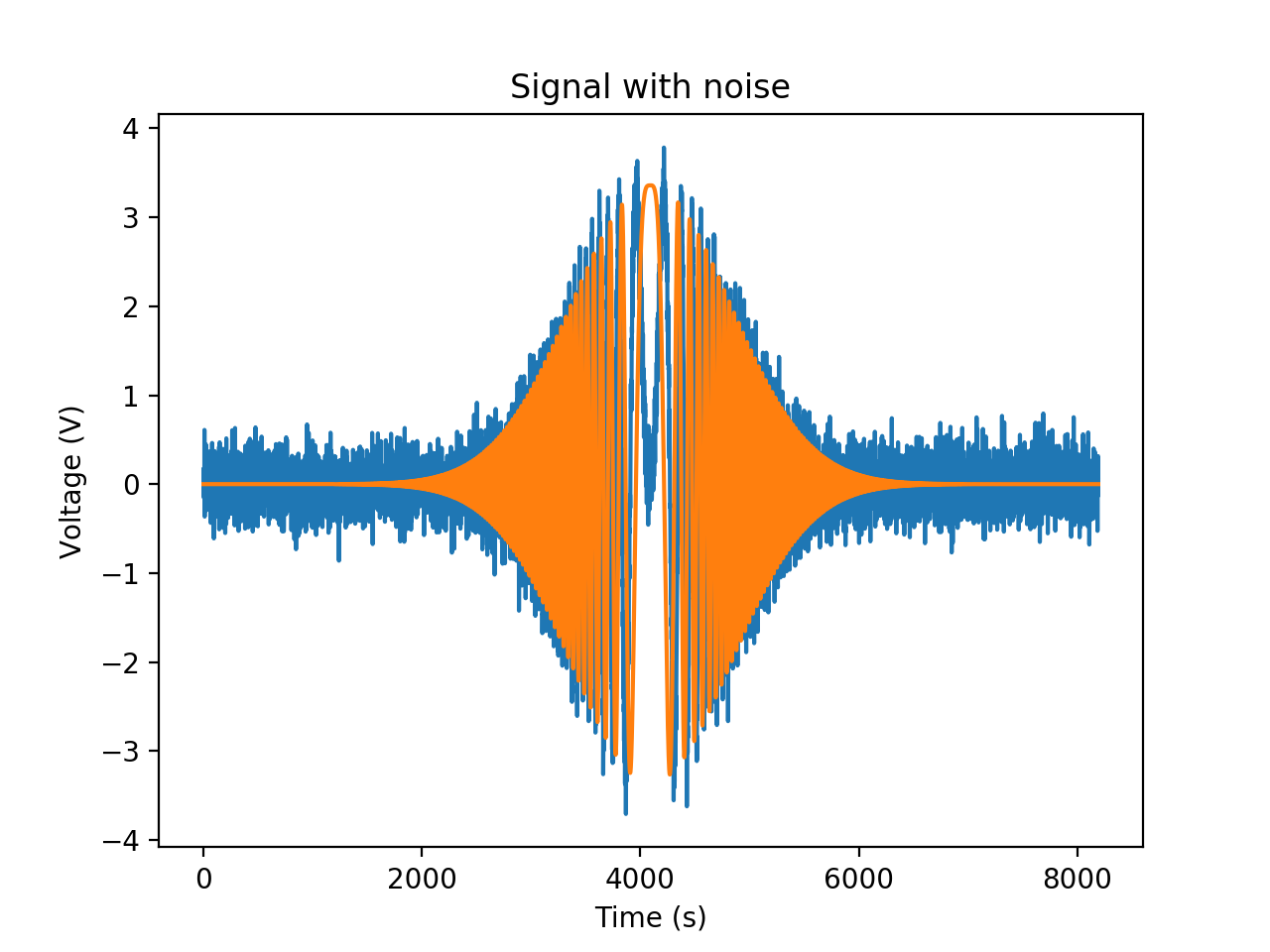}
            \caption{Chirping Sine-Gaussian}
            \begin{minipage}{.1cm}
            \vfill
            \end{minipage}
        \end{subfigure}
        \begin{subfigure}{\linewidth}
            \centering
            \includegraphics[width=3.5in]{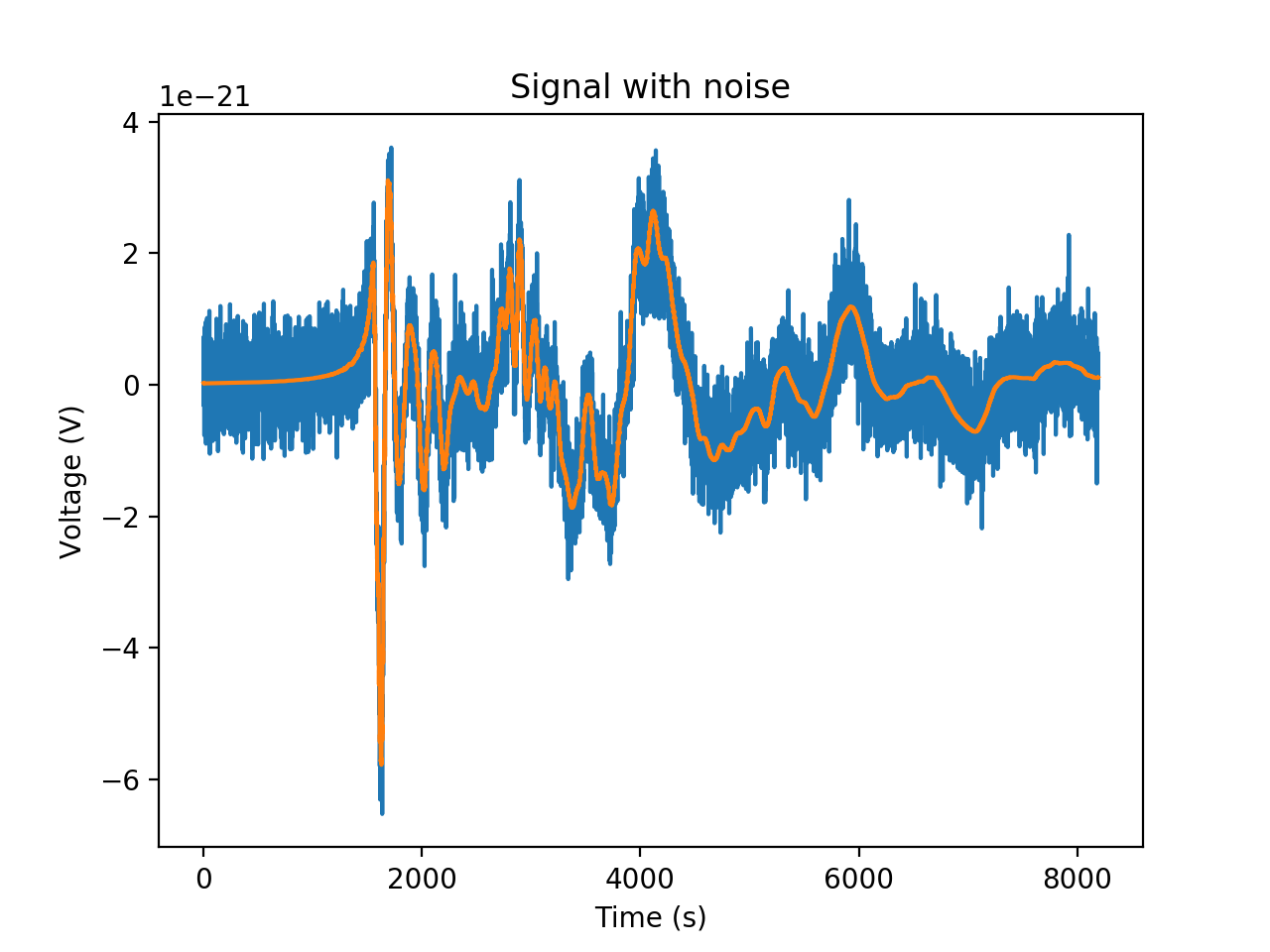}
            \caption{Supernova}
        \end{subfigure}
        \hfill
        \begin{subfigure}{\linewidth}
            \centering
            \includegraphics[width=3.5in]{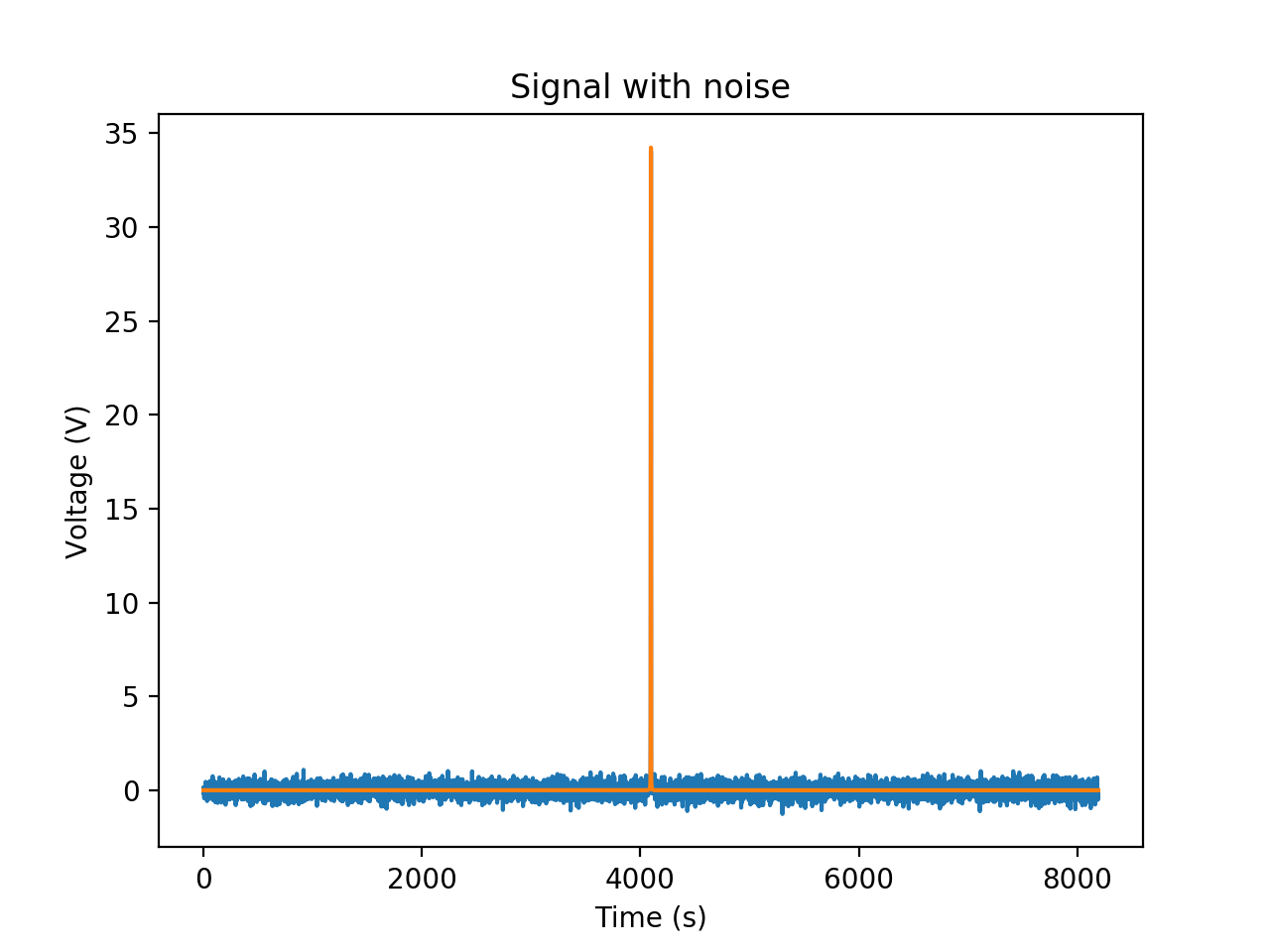}
            \caption{Cusp}
            \begin{minipage}{.1cm}
            \vfill
            \end{minipage}
        \end{subfigure}
    \phantomcaption
\end{figure}
\begin{figure}[H]
\ContinuedFloat
        \begin{subfigure}{\linewidth}
            \centering
            \includegraphics[width=3.5in]{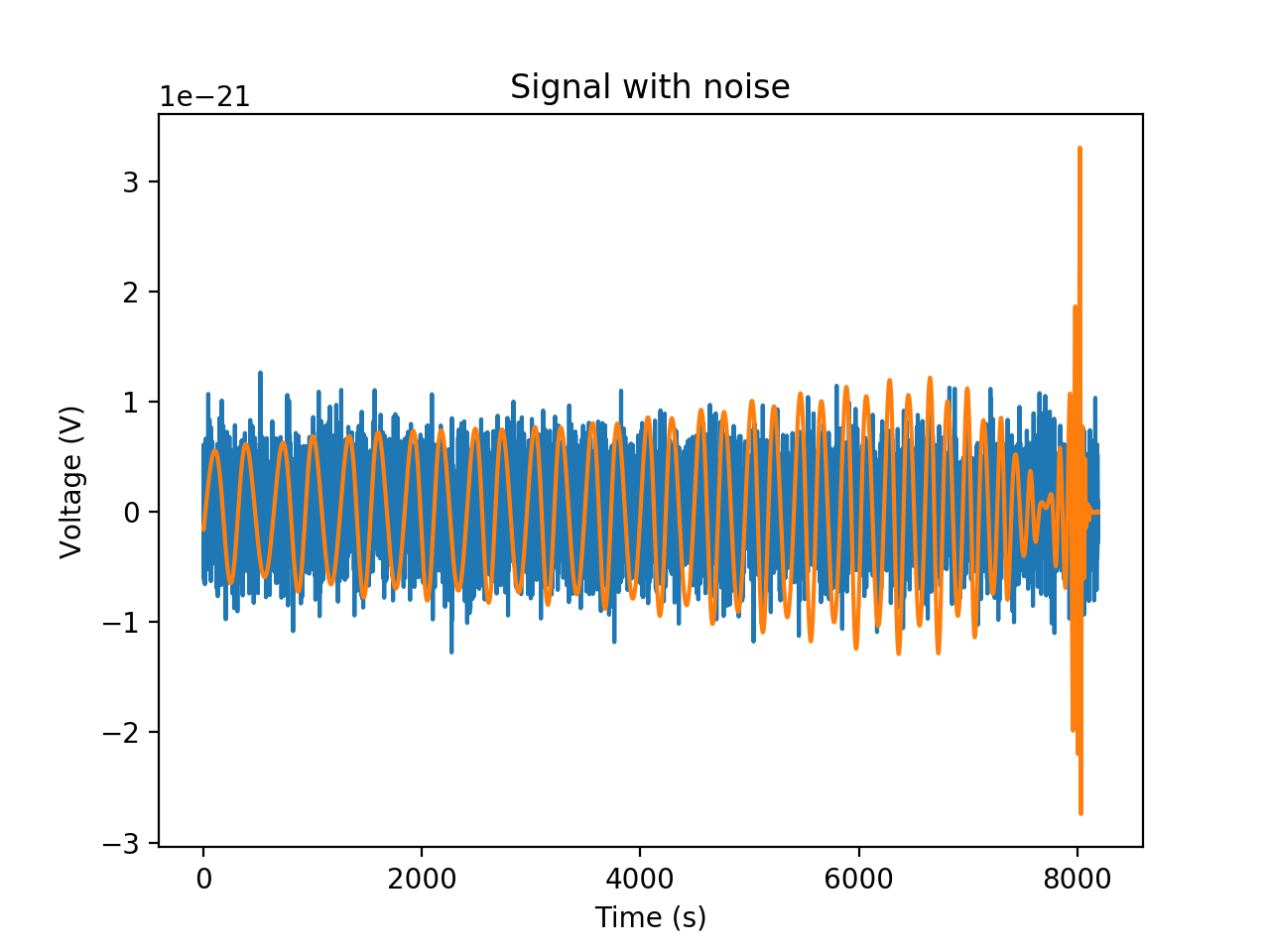}
            \caption{Black Hole Merger}
            \begin{minipage}{.1cm}
            \vfill
            \end{minipage}
        \end{subfigure}
        \begin{subfigure}{\linewidth}
            \centering
            \includegraphics[width=3.5in]{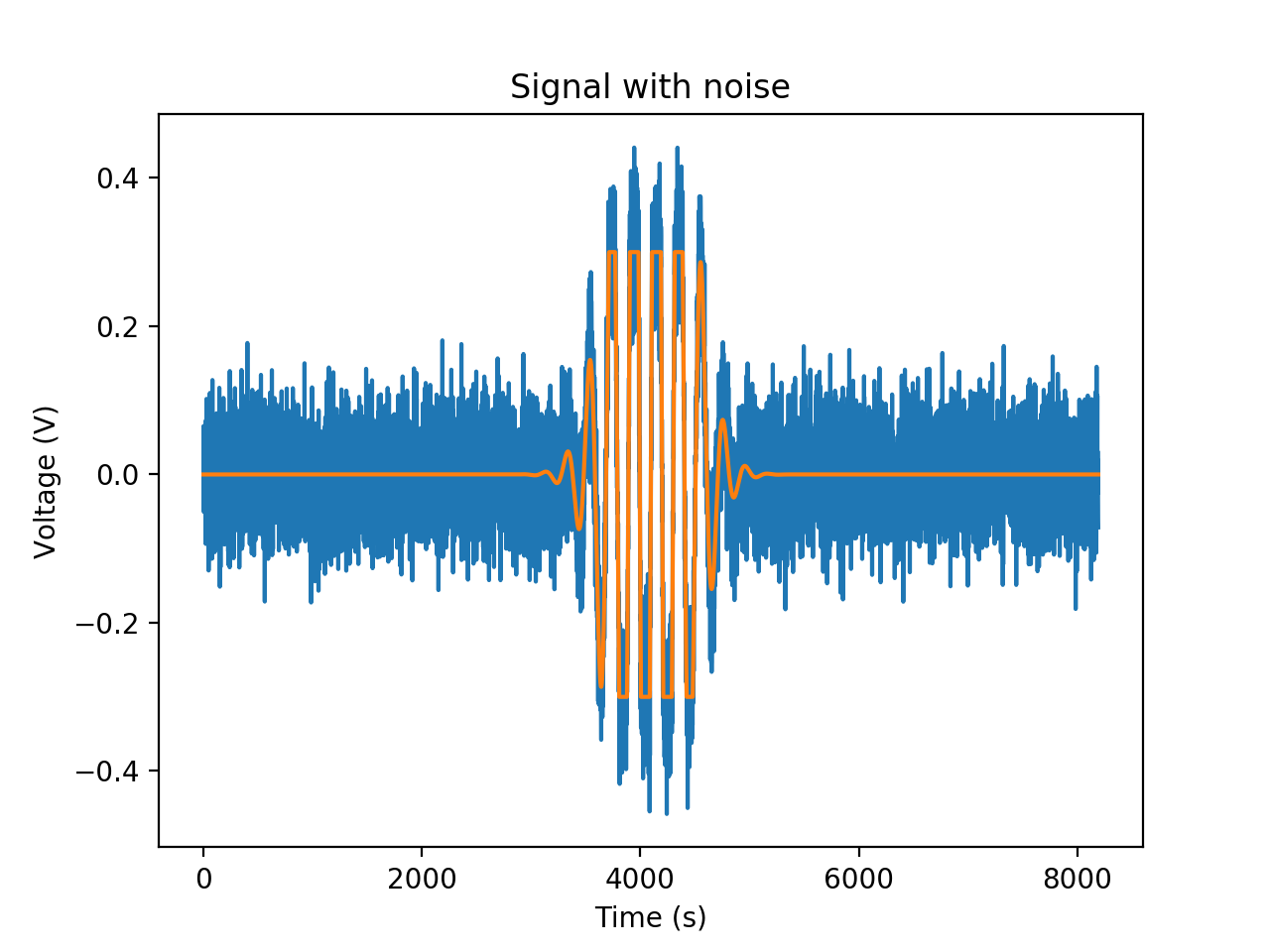}
            \caption{Blip}
            \begin{minipage}{.1cm}
            \vfill
            \end{minipage}
        \end{subfigure}
        \caption{Sample signals of the eight different types of transients. In order to find out whether there is noise present in the data, we use residual, i.e., we subtract the model data from experimental data and find out if this residual can fit the properties to a certain type of noise.}
\end{figure}

\section{Analysis}
In this project, we will be exploring the use of deep learning in classifying the various transient signals mentioned in the above section. Later, we will be using an unsupervised model such as Total variation, Principal Component Analysis, Support Vector Machine,  Wavelet decomposition or Random Forests on our generated dataset for feature extraction which will denoise the dataset and help our supervised model in better classifying the inputs

We will be training a CNN ( Convolutional Neural Network ) model and a Bidirectional LSTM-RNN ( An implementation of Recurrent Neural Network using  Long short-term memory ( LSTM ) modules ) model for time-series classification. We will be comparing the results of both models using standard metrics like Accuracy. Recurrent neural networks are widely used in various machine learning problems such as speech recognition, financial data analysis, time series forecasting, and time series classification.  The main feature that makes RNN highly versatile when working with time series data is their ability to learn sequences in the data. CNN is primarily used for image and video data but we can use a deep CNN model for classifying the transients due to its ability to learn non-linear hierarchical structures and lower computational costs. We will also explore the results obtained by the Deep Filtering algorithm proposed by Daniel George and E. A. Huerta. Apart from these, we will also be looking at other machine learning and deep learning algorithms such as stacked autoencoder, multilayer perceptron, random forests, logistic regression classifier , Support Vector Machine ( SVM ), and extreme learning machine ( ELM ). 

The architectures of the Deep Filtering algorithm, RNN, CNN, MLP, and stacked autoencoder have been presented in the figures below. For random forests, we have used 100 trees with the depth of each tree being 5. The ELM has been applied with 8000 hidden neurons and a hyperbolic tangent activation function. For implementing SVM, we have used LinearSVC ( Linear Support Vector Classification ) which uses “one-vs-the-rest” multi-class strategy. For training the gradient descent based algorithms we have used ADAM optimizer with a learning rate and decay of \( 10^{-5} \). CNN, RNN ,and Deep Filtering algorithm have been trained for 50 epochs with a batch size of 128 whereas the MLP and stacked autoencoder have been trained for 100 epochs with a batch size of 128. We have summarized the accuracy of trained models and the computational time taken for training the models on both a CPU ( i7 8750h ) and a GPU ( Tesla K80 ) respectively. We have used hold-out cross-validation for testing the performance of the models with a train-test split of 0.8-0.2 .  
\begin{figure}[H]
    \centering
    \includegraphics[width=1.7in]{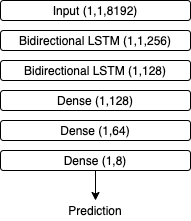}
    \caption{Architecture of LSTM Model}
\end{figure}

\begin{figure}[H]
    \centering
    \includegraphics[width=1.7in]{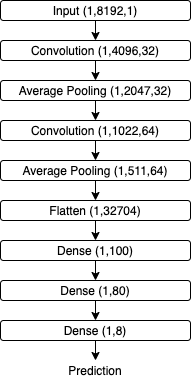}
    \caption{Architecture of CNN Model}
\end{figure}

\begin{figure}[H]
    \centering
    \includegraphics[width=1.7in]{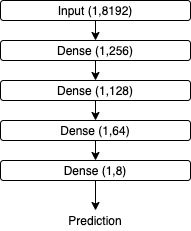}
    \caption{Architecture of MLP Model}
\end{figure}

\begin{figure}[H]
    \centering
    \includegraphics[width=1.7in]{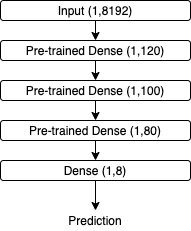}
    \caption{Architecture of Stacked Autoencoder Model}
\end{figure}

\begin{figure}[H]
    \centering
    \includegraphics[width=1.7in]{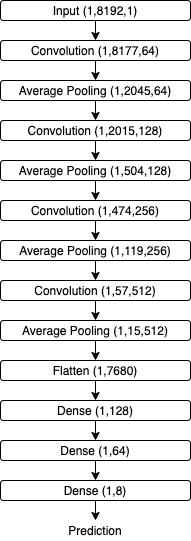}
    \caption{Architecture of Deep Filtering Model}
\end{figure}

\section{Results}
In this section, we will be presenting the predictions of each model  in the form of confusion matrices and also the accuracy values and the training times of each model in tables below.

\begin{figure}[H]
    \centering
    \includegraphics[width=3in]{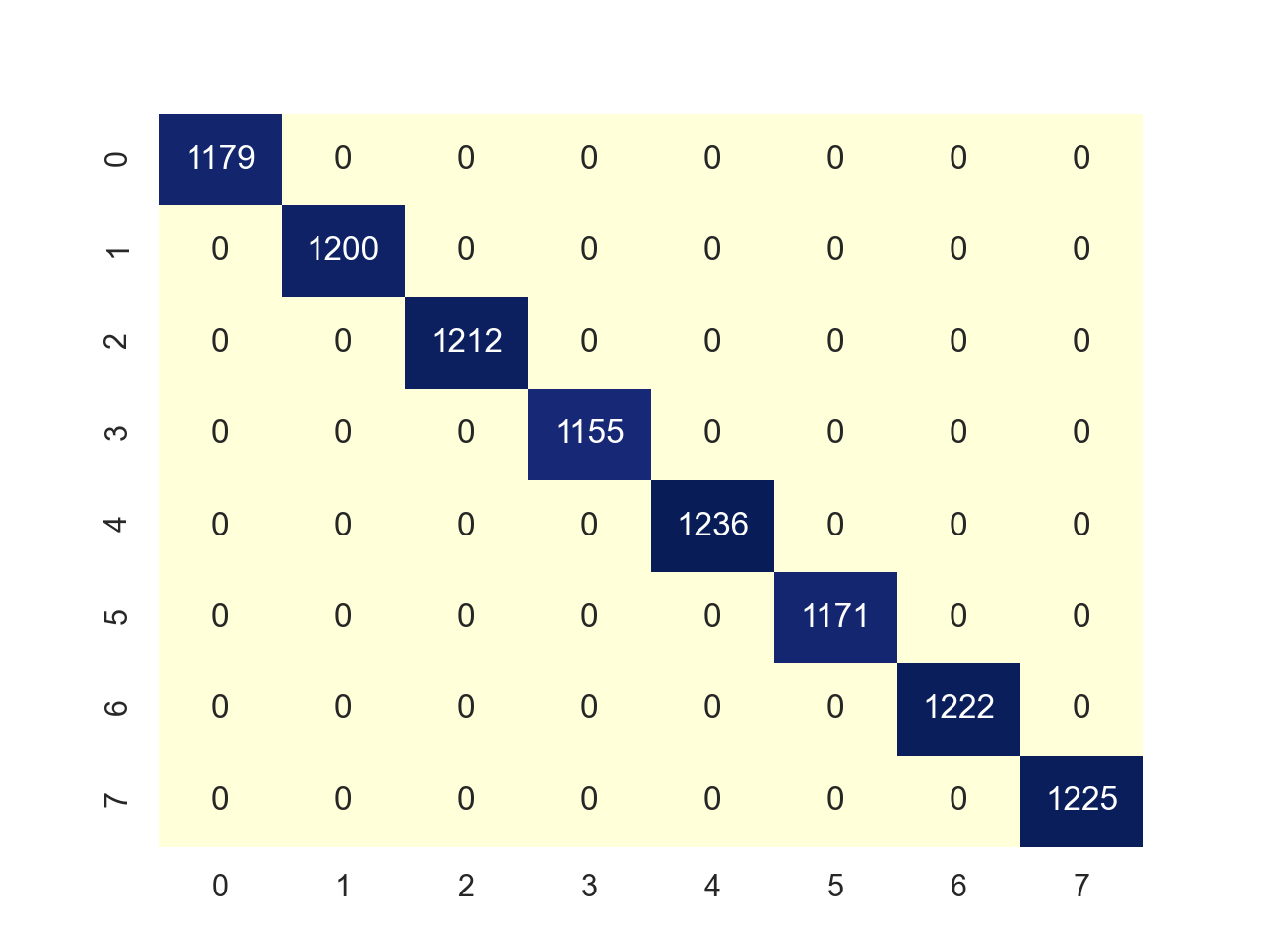}
    \caption{Prediction of CNN Model}
\end{figure}

\begin{figure}[H]
    \centering
    \includegraphics[width=3.2in]{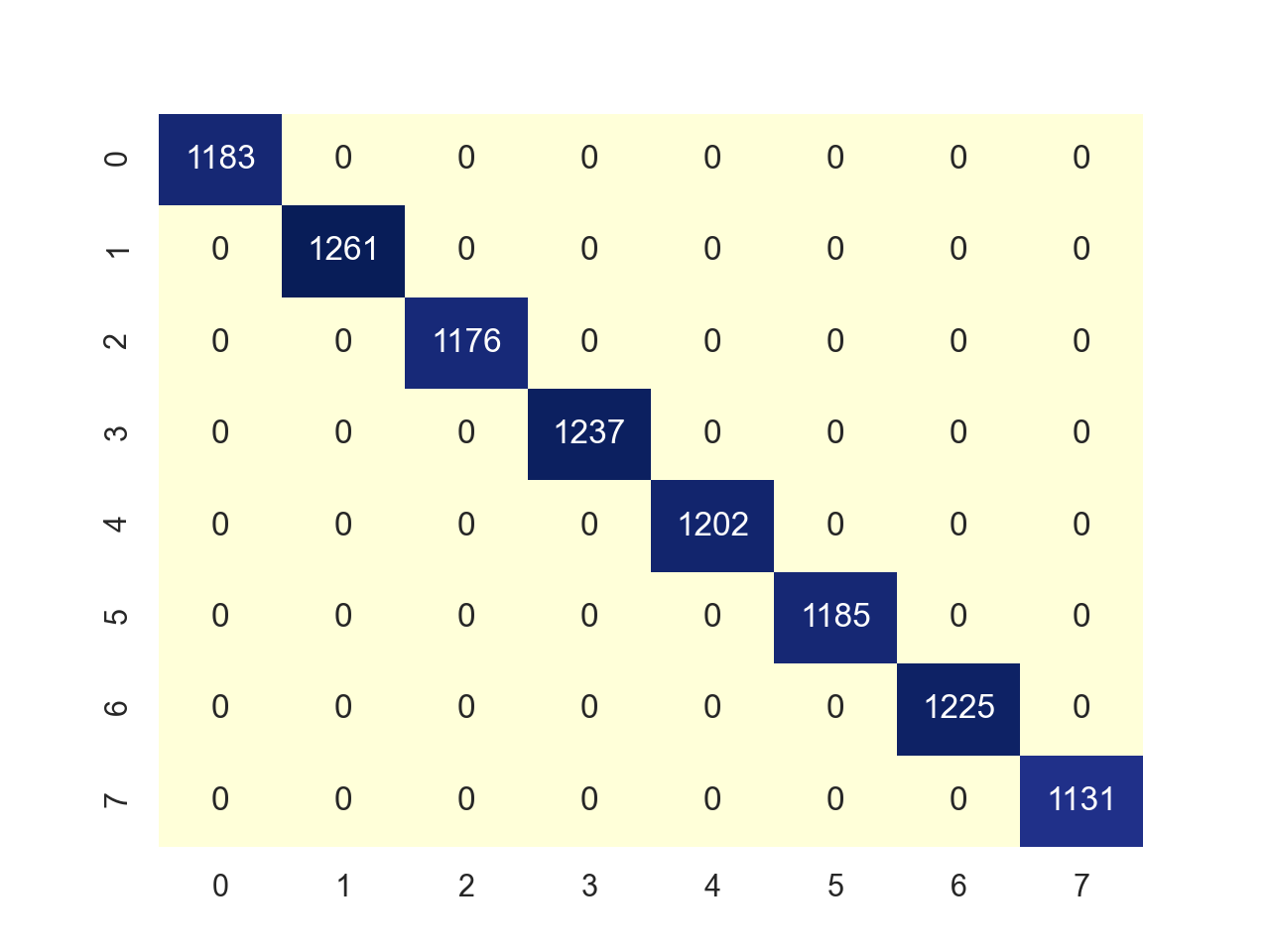}
    \caption{Prediction of Deep Filtering Model}
\end{figure}

\begin{figure}[H]
    \centering
    \includegraphics[width=3.2in]{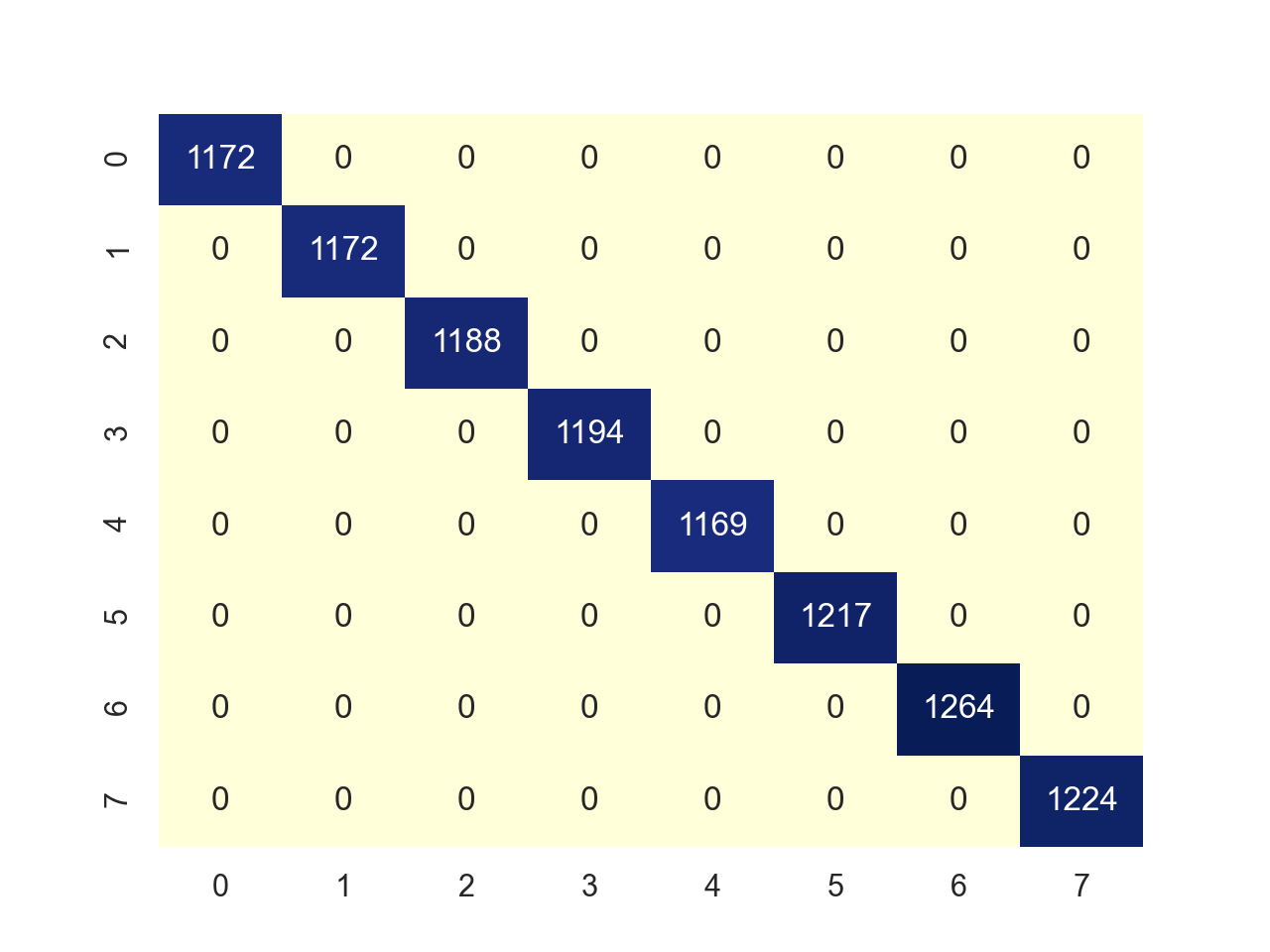}
    \caption{Prediction of RNN Model}
\end{figure}

\begin{figure}[H]
    \centering
    \includegraphics[width=3.2in]{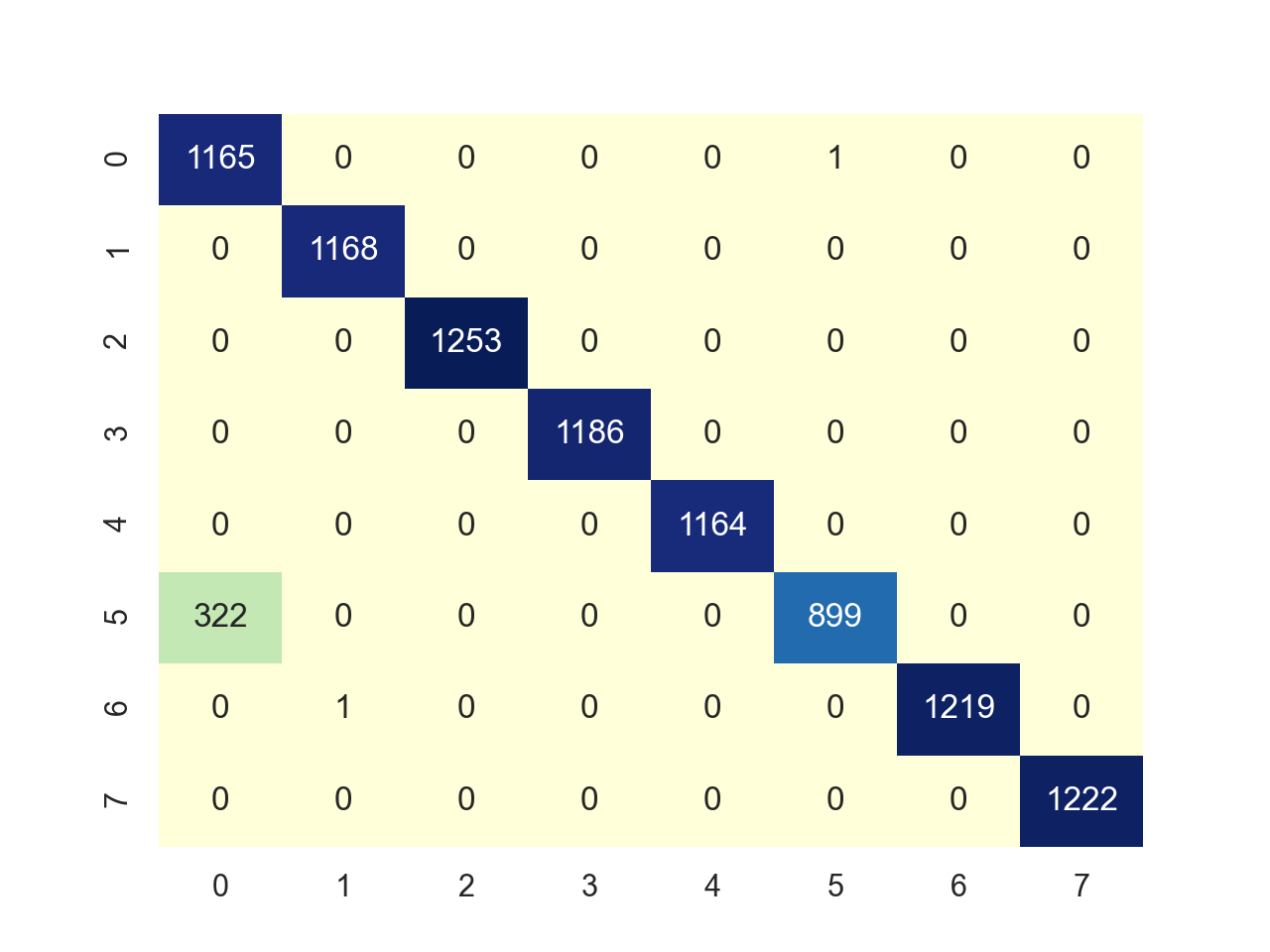}
    \caption{Prediction of MLP Model}
\end{figure}

\begin{figure}[H]
    \centering
    \includegraphics[width=3.2in]{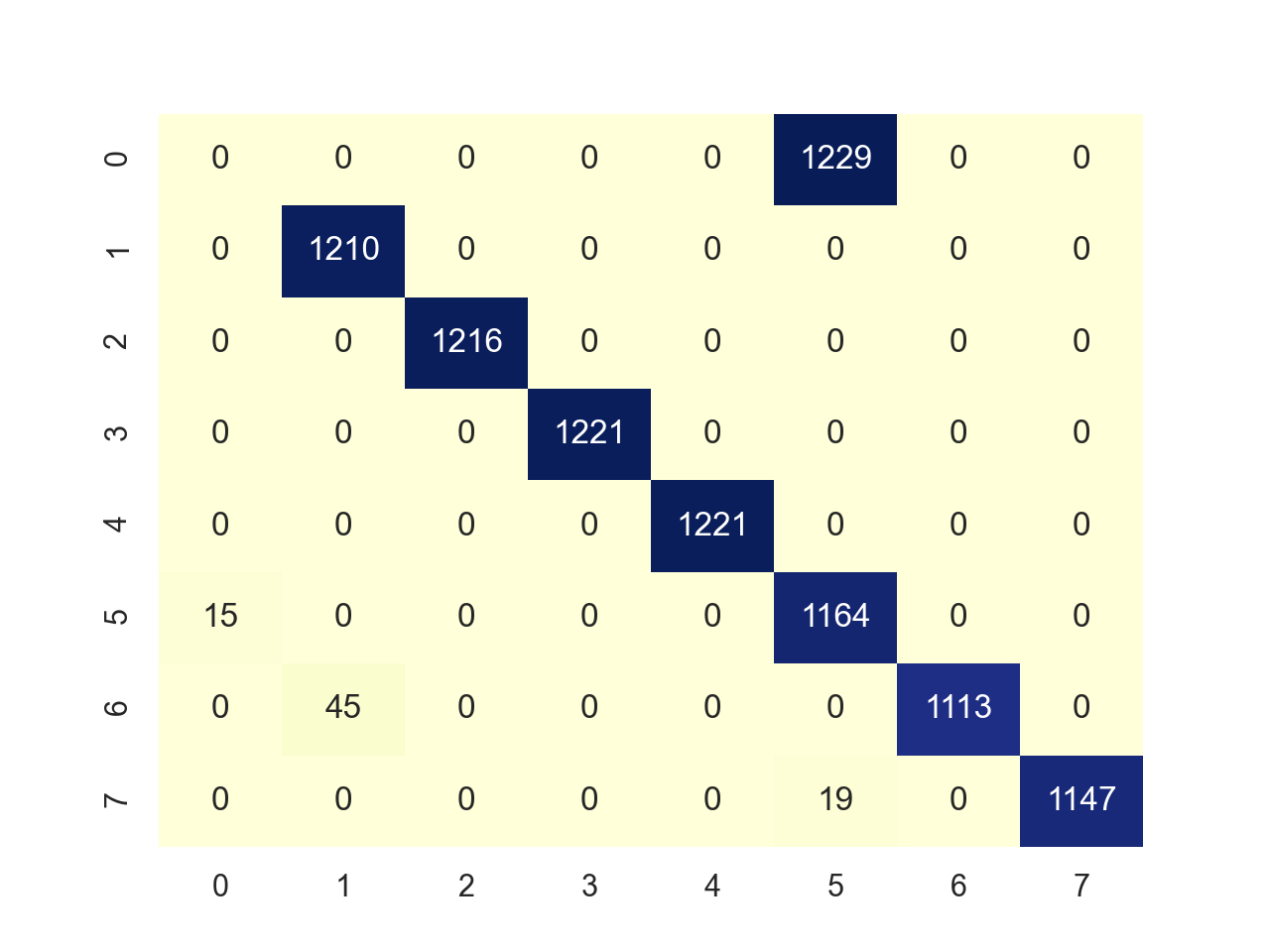}
    \caption{Prediction of Stacked Autoencoder Model}
\end{figure}

\begin{figure}[H]
    \centering
    \includegraphics[width=3.2in]{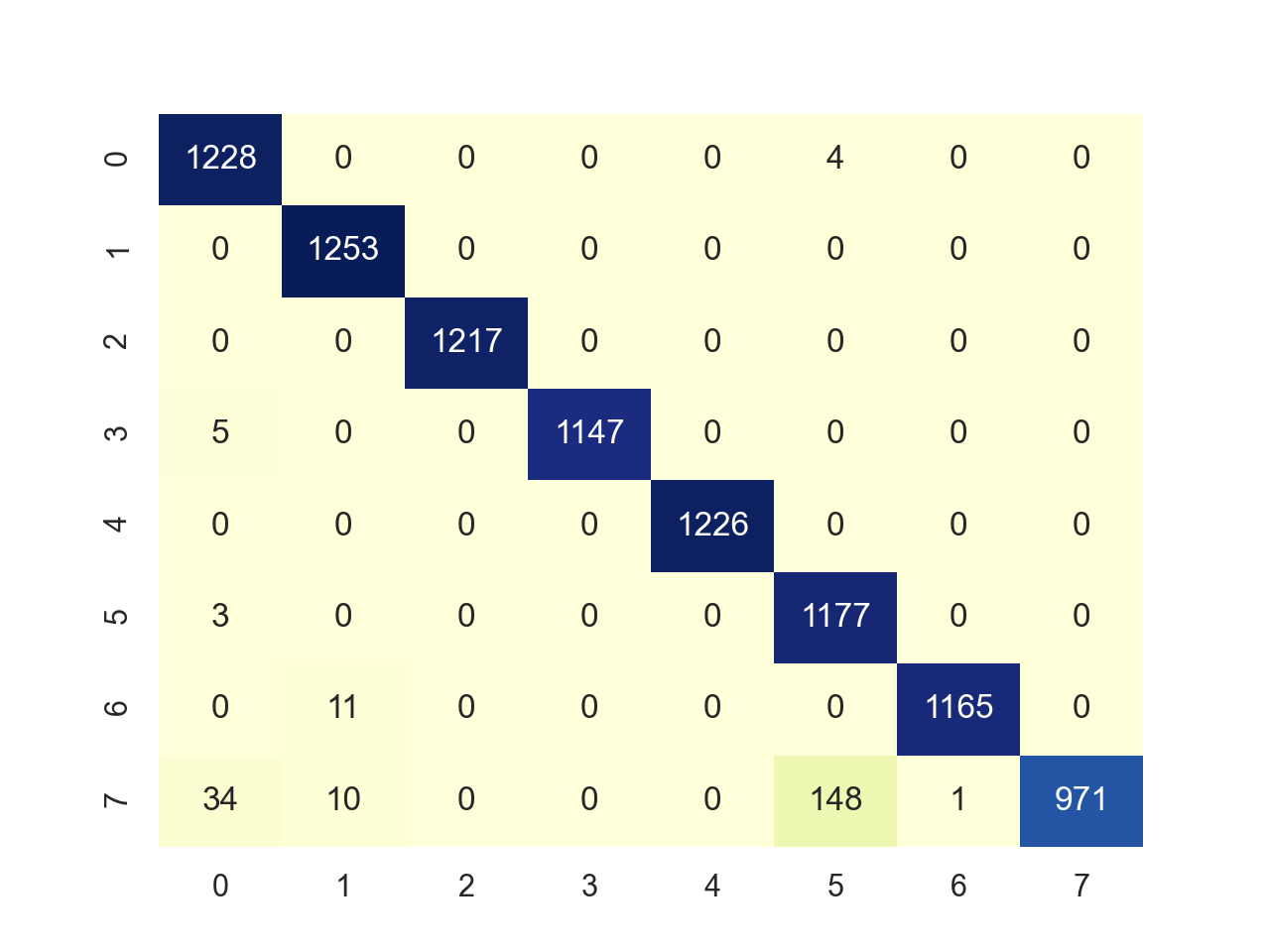}
    \caption{Prediction of Random Forest Model}
\end{figure}

\begin{figure}[H]
    \centering
    \includegraphics[width=3.2in]{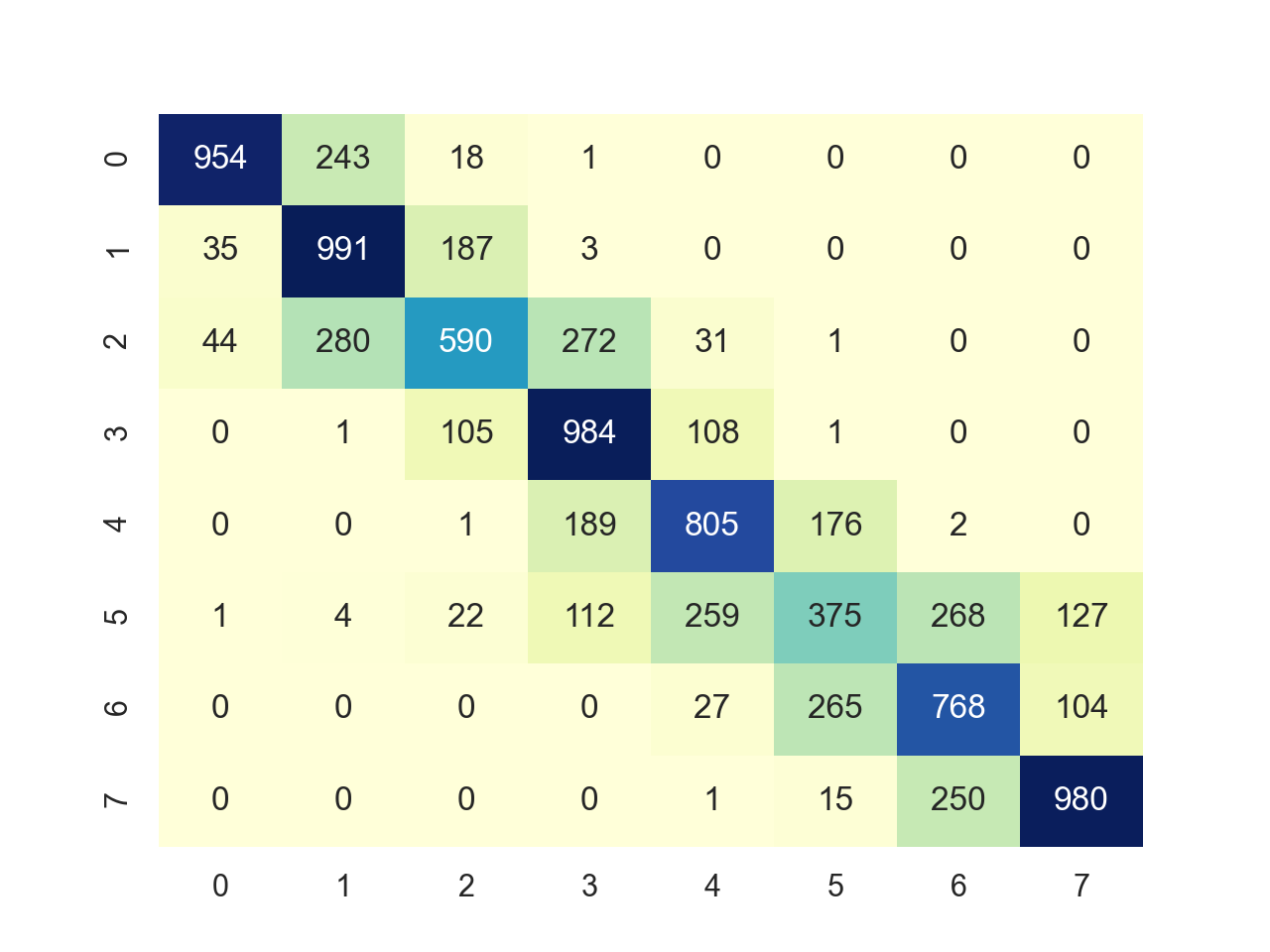}
    \caption{Prediction of ELM Model}
\end{figure}

\begin{figure}[H]
    \centering
    \includegraphics[width=3.2in]{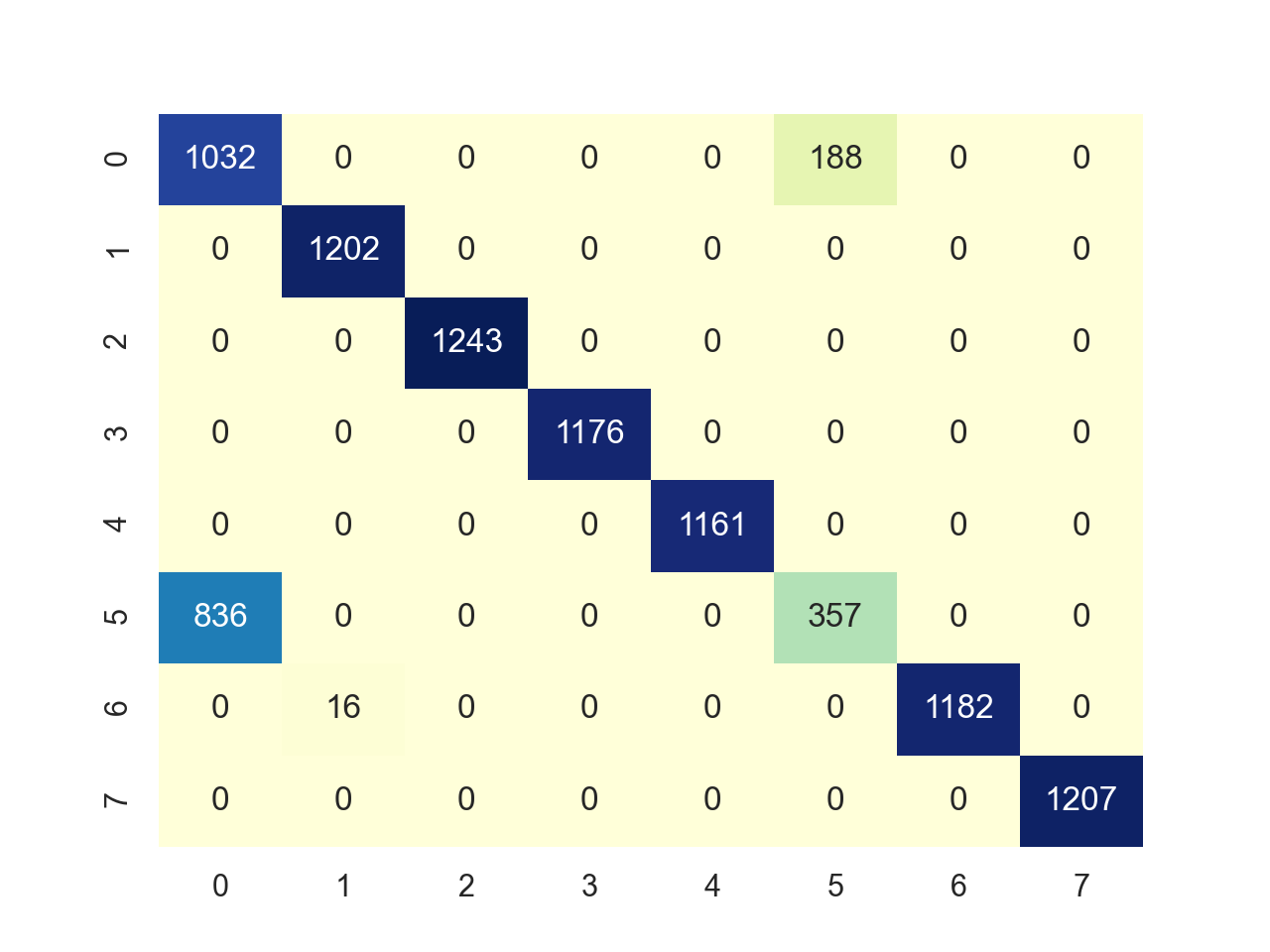}
    \caption{Prediction of Logistic Regression Classifier}
\end{figure}

\begin{figure}[H]
    \centering
    \includegraphics[width=3.2in]{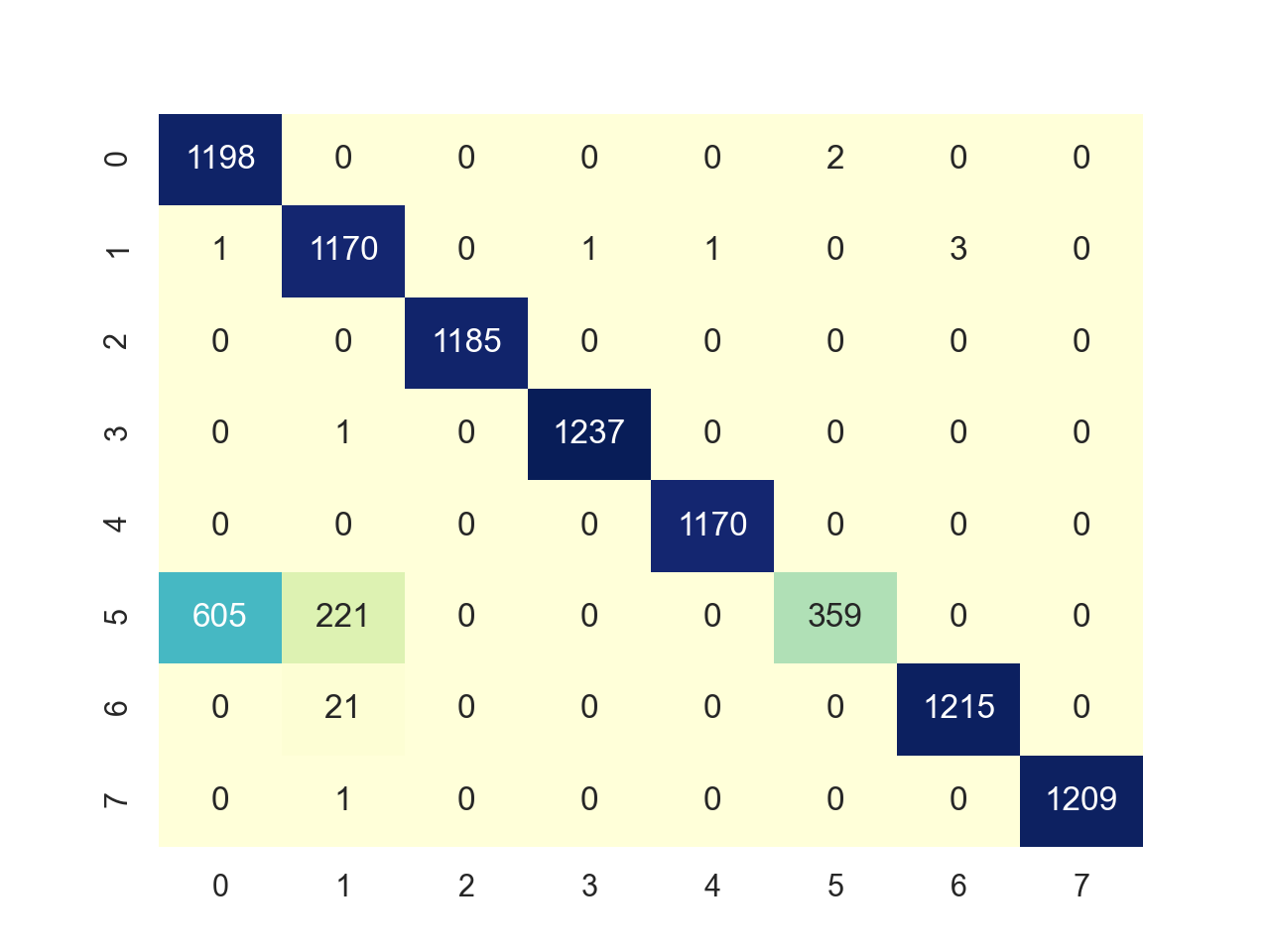}
    \caption{Prediction of SVM}
\end{figure}

\begin{table}[H]
\caption{Index corresponding to each transient}
\begin{tabular}{| >{\centering\arraybackslash}m{0.7in} | >{\centering\arraybackslash}m{2.5in} |}
\hline
\textbf{Index} & \textbf{Transient}       \\ \hline
0              & Ring Gaussian            \\ \hline
1              & Sine Gaussian            \\ \hline
2              & Gaussian                 \\ \hline
3              & Chirping Sine Gaussian   \\ \hline
4              & Cusp                     \\ \hline
5              & Binary Black Hole Merger \\ \hline
6              & Blip                     \\ \hline
7              & Supernova                \\ \hline
\end{tabular}
\end{table}

\begin{table}[H]
\caption{Accuracy of each model based on predictions from test data}
\begin{tabular}{| >{\centering\arraybackslash}m{2in} | >{\centering\arraybackslash}m{1.2in} |}
\hline
\textbf{Model}      & \textbf{Accuracy} \\ \hline
Deep Filtering      & 100.00000         \\ \hline
CNN                 & 100.00000         \\ \hline
RNN                 & 100.00000         \\ \hline
Random Forest       & 97.75100          \\ \hline
MLP                 & 96.62500          \\ \hline
SVM                 & 91.07291          \\ \hline
Logistic Regression & 89.66666          \\ \hline
Stacked Autoencoder & 86.37500          \\ \hline
ELM                 & 67.15625          \\ \hline
\end{tabular}
\end{table}

\begin{table}[H]
\caption{Computational time for training each model ( in seconds )}
\label{tab:my-table}
\begin{tabular}{| >{\centering\arraybackslash}m{2in} | >{\centering\arraybackslash}m{0.6in} | >{\centering\arraybackslash}m{0.6in} |}
\hline
\textbf{Model}      & \textbf{CPU} & \textbf{GPU} \\ \hline
Deep Filtering      & 73700        & 4800         \\ \hline
CNN                 & 4050         & 300          \\ \hline
RNN                 & 2200         & 200          \\ \hline
Stacked Autoencoder & \textit{650} & 520          \\ \hline
MLP                 & 600          & 300          \\ \hline
SVM                 & 1312         & -            \\ \hline
ELM                 & 379          & -            \\ \hline
Random Forest       & 86           & -            \\ \hline
Logistic Regression & 44           & -            \\ \hline
\end{tabular}
\end{table}

\section{Conclusion}
We have explored various machine learning and deep learning algorithms for the classification of transient signals. The convolution and lstm based models, Deep Filtering, CNN and RNN were able to achieve an accuracy of 100 percent over the test dataset. Compared to Deep Filtering and CNN, the RNN model was able to get an accuracy of 100 percent in a much lower training time which makes it ideal for implementation on low powered devices for real-time transient detection and classification. On the other hand, models such as random forests and logistic regression were able to achieve a decent accuracy rate in an extremely low training time which makes them ideal for fast inference to get an insight over the gravitational wave data that is available. We plan to continue this study by introducing more types of transient waves and also add other types of low-frequency noises to the transients apart from white noise. We will also compare deep filtering algorithm with RNN for classifying low SNR binary merger signals and white noise, which is an important task while analysing highly noisy time series data streams.  Our next objective is to compute residual and compare with different noise signatures. We will also be working on an unsupervised deep learning algorithm to learn the representations of merger signals and classify them from noise samples.

\clearpage

\nocite{*}

\bibliography{ligoref}

\end{document}